# Robust $L_\infty$-induced deconvolution filtering for linear stochastic systems and its application to fault reconstruction

Mehrdad Tabarraie*

Private Research Laboratory, Opposite Elahi-Parast, Tabriz 51668-46897, Iran

*Corresponding author

*E-mail address:* mehrdad.tabarraie@gmail.com

**Tel/Fax**: (+98) 411 3818953

**Abstract**

The problem of stationary robust $L_\infty$-induced deconvolution filtering for the uncertain continuous-time linear stochastic systems is addressed. The state space model of the system contains state- and input-dependent noise and deterministic parameter uncertainties residing in a given polytope. In the presence of input-dependent noise, we extend the derived lemma in Berman and Shaked (2010) characterizing the induced $L_\infty$ norm by linear matrix inequalities (LMIs), according to which we solve the deconvolution problem in the quadratic framework. By decoupling product terms between the Lyapunov matrix and system matrices, an improved version of the proposed $L_\infty$-induced norm bound lemma for continuous-time stochastic systems is obtained, which allows us to realize exploit parameter-dependent stability idea in the deconvolution filter design. The theories presented are utilized for sensor fault reconstruction in uncertain linear stochastic systems. The effectiveness and advantages of the proposed design methods are shown via two numerical examples.

**Keywords:** $L_\infty$-induced deconvolution filtering; linear stochastic system; polytopic uncertainty; parameter-dependent stability; fault reconstruction

## 1. Introduction

The aim of deconvolution filter is the estimation of the unknown input signal of a system by means of measured outputs. Deconvolution problem has comprehensive use in environments such as equalization, image restoration, speech processing, and fault detection (see [1,2] and the references therein). One of the widely used approaches for the deconvolution problem is the approach of minimum variance deconvolution, see [1,2] for a concise description. However, such approach has some limitations due to the fact that it requires the exact knowledge of the dynamic model of the system and the statistics of the noise signals. Therefore, its performance deteriorates significantly in presence of parameter uncertainties.

Two types of uncertainties have been considered in the literature. The first type is the deterministic uncertainties which are commonly posed in two forms: norm-bounded uncertainty and convex-polytopic uncertainty. Polytopic uncertainty is utilized exhaustively in robust estimation of uncertain systems whose parameters are not known thoroughly and lie in a given convex-bounded polytopic domain [1]. The second type is the stochastic uncertainties which have been considered to be multiplicative noise or Markov jump perturbations. Markov jump systems are efficiently used to model the systems which sudden variations occur in their structures. Such systems encountered in many practical applications, see e.g., [3,4]. In the case of stochastic systems with multiplicative noise, the parameter uncertainties are modeled as white noise processes [5]. Such models of uncertainties are common in many branches of science, e.g., economics, population models, immunology (see e.g., [5,6] and the references therein), and several engineering applications such as communication channel equalization, image processing, space navigation, and fault detection (see [7,8]).

In the recent years, the problems of $H_\infty$ and $L_2 - L_\infty$ deconvolution [9,10] have attracted attention for linear stochastic systems governed by the *Itô*-type stochastic differential equations with multiplicative white noise. The variance of the multiplicative noise depends on the states or inputs of the system. In contrast with the minimum variance deconvolution approach, $H_\infty$ and $L_2 - L_\infty$ methods require no statistical knowledge about the exogenous disturbances, and the disturbances should only have bounded energy. Likewise, these filters are robust against parameter uncertainties. In [11], a stochastic bounded real lemma was presented in terms of LMIs, which has played an important role in $H_\infty$ filtering design for the linear continues-time stochastic systems. Based on this important lemma, a proper filter has been proposed for the stochastic reduced-order $H_\infty$ estimating in [12] which estimates a linear combination of state and exogenous input signals, and in [9] an $H_\infty$ deconvolution filter has also been designed for the linear stochastic systems with state-multiplicative (state-dependent) noise and deterministic interval uncertainties. In the discrete-time case, deconvolution filtering for communication channels in which the modeling error is characterized in terms of multiplicative white noise perturbations has been designed by minimax approaches in state space setup in [13] and via a polynomial approach in [14].

Although in designing $H_\infty$ and $L_2 - L_\infty$ filters the exogenous disturbances are supposedly of $L_2$ type (energy-bounded signals), in practice, they often have a bounded peak (of $L_\infty$ type). These disturbances are known as persistent bounded disturbances which are studied in the $L_1$ control theory [15]. For such types of disturbances, the induced operator norm is the induced $L_\infty$ or 'peak to peak' norm of the linear system under investigation (or the $L_1$ norm of its impulse response). Based on the approach of bounding the reachable set with inescapable ellipsoids and the LMI technique, peak-to-peak gain minimization problem of linear continuous-time systems has been solved in [16]. In this problem, the objective is minimizing the ∗-norm, which is an upper bound on the induced $L_\infty$ norm, rather than minimizing the induced $L_\infty$ norm directly. In a manner similar to [16], Scherer et al. [17] have presented a sufficient condition in terms of LMIs for an upper bound of the peak-to-peak norm of the underlying system. In discrete-time counterpart, results similar to [17] have been obtained in [18,19]. The problem of designing robust filter for uncertain polytopic linear systems with guaranteed $L_1$ performance has been addressed in [20,21] by using the results of [17,18], respectively. In the case of state-multiplicative stochastic systems, recently Berman and Shaked [22] have extended the cited condition in [17] and established an important lemma according to which stochastic estimation in the induced $L_\infty$ norm sense is discussed. Motivated by the ideas on [16], a robust peak-to-peak filter recently has also been designed for a class of uncertain Markov jump systems in [4].

The aforementioned robust peak-to-peak filtering methods are based on the notion of quadratic stability. These methods can produce conservative results, because the same parameter-independent Lyapunov function must be used for the entire uncertainty domain [23]. To reduce the conservatism of the quadratic framework, the parameter-dependent stability idea has been utilized by some researchers to investigate the $H_\infty$ and $L_2 - L_\infty$ filtering problems for uncertain polytopic linear systems, see e.g., [23,24]. In [25,26] less-conservative $L_1$ performance conditions in terms of LMIs have been proposed by decoupling the product terms between the Lyapunov matrix and system matrices by introducing slack matrix variables, upon which the robust peak-to-peak filters are obtained from the solution of convex optimization problems. In the stochastic context, two parameter-dependent continuous-time bounded real lemmas are proposed in [27] for uncertain polytopic linear systems with stochastic uncertainties. By using the parameter-dependent Lyapunov function method, the problem of $H_\infty$ filtering for linear stochastic systems with polytopic uncertainties has been investigated in [5] for the continuous-time case and in [28] for the discrete-time case. Recently the robust $H_\infty$ and $L_2 - L_\infty$ deconvolution filters have been designed in [10] for uncertain linear stochastic systems by using homogeneous polynomial parameter-dependent matrix (HPPDM) approach.

We extend the results of the stochastic peak-to-peak filtering [22] to the deconvolution problem for linear stochastic systems, in the stationary case. In this paper, we propose two approaches, namely robust stochastic $L_\infty$-induced deconvolution filtering and improved robust stochastic $L_\infty$-induced deconvolution filtering. At the first we extend lemma 1 of [22] to the case where input-dependent noise is considered. Based on this developed lemma, the induced $L_\infty$ deconvolution filtering is designed using LMI techniques in the quadratic framework. The proposed deconvolution filter is also extended to the case in which the deterministic component of state space model matrices and the covariance matrices of multiplicative noises are uncertain and reside in a given polytopic type domain. By decoupling product terms between the Lyapunov matrix and system matrices, an improved version of the proposed $L_\infty$-induced norm bound Lemma for continuous-time stochastic systems is obtained, which allows us to realize exploit parameter-dependent stability idea in the deconvolution filter design. The obtained simulation results for parameter-dependent Lyapunov approach are less conservative than the existing one in the quadratic framework. We then solve the sensor fault reconstruction problem with following the lines of [29] by a deconvolution filter in the induced $L_\infty$ norm setting and demonstrate the applicability of the results regarding the fault estimation, via an inverted pendulum system with multiplicative noise.

**Notation.** The superscript $T$ shows matrix transposition. $\Re^n$ determines the $n$-dimensional Euclidean space, and $\|\cdot\|$ is the Euclidean vector norm, and $\Re^{n\times m}$ is a set of all the $n\times m$ real matrices. The notation $P>0$ for $P^{n\times n}$ means that $P$ is symmetric and positive definite. $E\{\cdot\}$ stands for expectation. The symbol $*$ is used for the symmetric terms in a symmetric matrix. $I_n$ is the $n\times n$ identity matrix. By $L_\vartheta(\Re^k)$ we denote the space of bounded $\Re^k$-valued functions on the probability space $(\Omega,\vartheta,\Psi)$, where $\Omega$ is the sample space, $\vartheta$ is an $\sigma$-algebra of subsets of the sample space, and $\Psi$ is a probability measure on $\vartheta$. By $(\vartheta_t)_{t>0}$ we denote an increasing family of $\sigma$-algebras $\vartheta_t\subset\vartheta$. Likewise, let $L_{\vartheta_t}^\infty(\Re^k)$ denote the space of non-anticipative stochastic process $f(\cdot)=(f(t))_{t\in[0,\infty)}$ in $\Re^k$ with respect to $(\vartheta_t)_{t\in[0,\infty)}$ which satisfies $\|f\|_\infty \triangleq \sup_{t\geq 0}\sqrt{E\{\|f(t)\|^2\}} < \infty$. It should be mentioned that stochastic differential equations are of *Itô* type.

## 2. Problem formulation

We consider the following linear stochastic system with state- and input-dependent noise:

$$\begin{aligned}dx(t) &= [Ax(t)+B_1\omega(t)]dt + [G_1 x(t)+G_2\omega(t)]d\beta(t),\ x(0)=0,\\ y(t) &= C_2 x(t)+D_2\omega(t),\\ z(t) &= C_1 x(t)+D_{11}\omega(t)\end{aligned} \quad (1\text{a-c})$$

where $x\in\Re^n$ is the system state vector, and $\omega(t)\in L_{\vartheta_t}^\infty(\Re^q)$ is the exogenous input vector. $y\in\Re^r$ is the measurement vector, and $z\in\Re^m$ is the signal to be estimated. $A$, $B_1$, $C_1$, $C_2$, $D_{11}$, $D_2$, $G_1$, and $G_2$ are constant matrices with appropriate dimensions. $\beta(t)$ is a zero-mean real scalar Wiener process which satisfies $E\{d\beta(t)^2\}=dt$. In fact, $G_1\dot{\beta}$ and $G_2\dot{\beta}$ can be interpreted as white noise parameter perturbations in the matrices $A$ and $B_1$ respectively by adopting the fact that white noise signals are formally the derivatives of Wiener processes.

During this paper, we adopt the following definition.

**Definition 1** [30]. The system (1a) with $\omega(t)=0$ is called exponentially stable in mean square (ESMS) if there exist $\alpha>0$ and $\beta\geq 1$ such that $E\{\|x(t)\|^2\}\leq \beta e^{-\alpha t}\|x(0)\|^2$ for all $t\geq 0$ and $x(0)\in\Re^n$.

In the following theorem, necessary and sufficient conditions for exponential stability in the mean square sense are given.

**Proposition 1** [30]. The system (1a) is ESMS if and only if there exists $Q > 0$ such that $A^T Q + QA + G_1^T Q G_1 < 0$.

We take into account the following deconvolution filter to estimate $z(t)$:

$$d\hat{x} = A_f \hat{x} dt + B_f y dt, \quad \hat{x}(0) = 0,$$
$$\hat{z} = C_f \hat{x} + D_f y \quad (2)$$

where $\hat{x} \in \Re^n$ and $\hat{z} \in \Re^m$. Denoting $\tilde{z}(t) = z(t) - \hat{z}(t)$ and for a given scalar $\gamma > 0$, the following cost function is defined:

$$J_S \triangleq \|\tilde{z}\|_\infty - \gamma \|\omega\|_\infty \quad (3)$$

Considering the ESMS system of (1a-c) and the filter of (2) and denoting $\xi = \begin{bmatrix} x^T & \hat{x}^T \end{bmatrix}^T$, the following augmented system, which shows the filtering error dynamic, will be obtained:

$$d\xi = \tilde{A}\xi dt + \tilde{B}\omega dt + \left[\tilde{G}_1 \xi + \tilde{G}_2 \omega\right] d\beta, \quad \tilde{z} = \tilde{C}\xi + \tilde{D}\omega \quad (4)$$

where

$$\tilde{A} = \begin{bmatrix} A & 0 \\ B_f C_2 & A_f \end{bmatrix}, \quad \tilde{B} = \begin{bmatrix} B_1 \\ B_f D_2 \end{bmatrix}, \quad \tilde{G}_1 = \begin{bmatrix} G_1 & 0 \\ 0 & 0 \end{bmatrix},$$
$$\tilde{G}_2 = \begin{bmatrix} G_2 \\ 0 \end{bmatrix}, \quad \tilde{C} = \begin{bmatrix} C_1 - D_f C_2 & -C_f \end{bmatrix}, \quad \tilde{D} = D_{11} - D_f D_2 \quad (5)$$

Now, we assume that the matrices $A$, $B_1$, $C_1$, $C_2$, $D_{11}$, $D_2$, $G_1$, and $G_2$ are with partially unknown parameters. They reside in the polytope as follows:

$$\bar{\Omega} \triangleq \left\{ (A, B_1, C_1, C_2, D_{11}, D_2, G_1, G_2) \mid (A, B_1, C_1, C_2, D_{11}, D_2, G_1, G_2) = \sum_{i=1}^s \alpha_i \bar{\Omega}_i, \alpha_i \geq 0, \sum_{i=1}^s \alpha_i = 1 \right\} \quad (6)$$

where $\bar{\Omega}_i \triangleq (A_i, B_{1i}, C_{1i}, C_{2i}, D_{11,i}, D_{2i}, G_{1i}, G_{2i})$, $i = 1,...,s$ are the polytope vertices.

Given $\gamma > 0$, the aim of robust stochastic peak-to-peak deconvolution filter is to seek for estimation $\hat{z}(t)$ from $z(t)$ over the infinite time horizon $[0, \infty)$ in such a way that $J_S$ of (3) is negative for all nonzero $\omega(t) \in L_{\mathcal{F}_t}^\infty(\Re^q)$ and for all uncertainties belonging to the polytope (6), that is, $\|\tilde{z}\|_\infty^2 / \|\omega\|_\infty^2 < \gamma^2$ which implies that the ratio between the peak value of the mean-square of the estimation error, i.e., $\|\tilde{z}\|_\infty^2$ and the peak value of the mean-square of the exogenous inputs, namely, $\|\omega\|_\infty^2$ is bounded by the prescribed $\gamma^2$.

**Remark 1.** Since $z(t)$ includes both state and unknown input of the system, the filter of (2) can be used for estimating a linear combination of the state and of the input.

## 3. Main results

Using Proposition 1, we make the following lemma which is the extension of Lemma 1 of [22]. The Lemma 1 gives a sufficient condition for an upper bound of the peak-to-peak norm of the system (4).

**Lemma 1.** Consider the ESMS system (1a-c) without polytopic uncertainties and the deconvolution filter (2) and let $\gamma > 0$ be given. The filtering error system (4) is ESMS, and $J_S$ of (3) is negative for all nonzero $\omega(t) \in L^\infty_{\mathcal{F}_t}(\mathfrak{R}^q)$ if there exist $Q > 0$, $\lambda$, and $\mu > 0$ so that $\mu < \gamma$, satisfying the following two matrix inequalities:

$$\begin{bmatrix} \tilde{A}^T Q + Q\tilde{A} + \lambda Q & Q\tilde{B} & \tilde{G}_1^T Q \\ \tilde{B}^T Q & -\mu I_q & \tilde{G}_2^T Q \\ Q\tilde{G}_1 & Q\tilde{G}_2 & -Q \end{bmatrix} < 0, \tag{7}$$

$$\begin{bmatrix} \lambda Q & 0 & \tilde{C}^T \\ 0 & (\gamma - \mu)I_q & \tilde{D}^T \\ \tilde{C} & \tilde{D} & \gamma I_m \end{bmatrix} > 0 \tag{8}$$

where $\lambda \in \left(0, -2\max\left(real\left\{eig(\tilde{A})\right\}\right)\right)$.

**Proof.** See Appendix A.

**Remark 2.** Note that the inequalities (7) and (8) are nonlinear duo to products between $\lambda$ and $Q$. Therefore, (7) and (8) become LMIs only if the variable $\lambda$ is fixed.

**Remark 3.** It should be emphasized that Lemma 1 is based on mean-square quadratic stability notion. When the Lemma 1 is extended to cope with the stochastic system (1) with uncertainties residing in the polytope (6), the Lyapunov matrix $Q$ is set to be fixed, which is required to satisfy each vertex of the polytope (6). Therefore, such extended lemma may lead to very conservative results. Motivated by the approach used in [31], an improved LMI representation of stochastic bounded real lemma has been derived for uncertain continuous-time stochastic systems in [5]. Following the same lines used in [5], we will present an improved version of Lemma 1 by decoupling the product terms between the Lyapunov matrix $Q$ and system matrices.

**Lemma 2.** Consider the ESMS system (1a-c) without polytopic uncertainties and the deconvolution filter (2) and let $\gamma > 0$ be given. The filtering error system (4) is ESMS, and $J_S$ of (3) is negative for all nonzero $\omega(t) \in L^\infty_{\mathcal{F}_t}(\mathfrak{R}^q)$ if there exist $Q > 0$, $W$, a sufficiently small scalar $\varepsilon > 0$, $\lambda \in \left(0, -2\max\left(real\left\{eig(\tilde{A})\right\}\right)\right)$, and $\mu > 0$ so that $\mu < \gamma$, satisfying (8) and the following inequality:

$$\begin{bmatrix} Q-W-W^T & W^T\left((1+\lambda\varepsilon/2)I+\varepsilon\tilde{A}\right) & \sqrt{\varepsilon}W^T\tilde{B} & 0 \\ * & -Q & 0 & \sqrt{\varepsilon}\tilde{G}_1^T W \\ * & * & -\mu I & \tilde{G}_2^T W \\ * & * & * & Q-W-W^T \end{bmatrix} < 0 \qquad (9)$$

**Proof.** See Appendix B.

**Remark 4.** The advantage of Lemma 2 lies in the fact that by introducing the additional slack variable $W$ and a sufficient small positive constant $\varepsilon$, it eliminates the product terms between the Lyapunov matrix $Q$ and the system matrices. This feature enables to use parameter-dependent Lyapunov functions for the whole uncertain domain, i.e., to use a different positive definite matrix for each vertex of a polytope. Therefore, the new induced $L_\infty$ performance condition leads to less conservative results when used for analysis and synthesis of polytopic uncertain stochastic systems.

## 4. Robust stochastic $L_\infty$-induced deconvolution filter

In this section, we solve the robust stochastic $L_\infty$-induced deconvolution filtering problem based on the lemma 1 in the quadratic framework. To facilitate the presentation of our results, first consider the stochastic system (1) is without polytopic uncertainties. Now, we give the following theorem.

**Theorem 1.** We consider the ESMS system (1a-c) without polytopic uncertainties and the deconvolution filter of (2), and let $\gamma > 0$ and $\lambda > 0$ be given constants. Then the following holds:

(a) The filtering error system (4) is ESMS, and $J_S$ of (3) is negative for all nonzero $\omega(t) \in L_{\mathcal{F}_t}^\infty(\mathfrak{R}^q)$, if there exist $R = R^T \in \mathfrak{R}^{n\times n}$, $V = V^T \in \mathfrak{R}^{n\times n}$, $Z \in \mathfrak{R}^{n\times r}$, $S \in \mathfrak{R}^{n\times n}$, $T \in \mathfrak{R}^{m\times n}$, $D_f \in \mathfrak{R}^{m\times r}$, and positive scalar $\mu$ so that $\mu < \gamma$, such that the following two LMIs are satisfied:

$$\Sigma_1(R,V,Z,S,\mu) < 0,$$
$$\Sigma_2(R,V,T,D_f,\mu) > 0, \qquad (10a,b)$$

where

$$\Sigma_1 \triangleq \begin{bmatrix} RA+A^TR+\lambda R & A^TV+C_2^TZ^T+S^T & RB_1 & G_1^TR & G_1^TV \\ * & -S-S^T+\lambda V & VB_1+ZD_2 & 0 & 0 \\ * & * & -\mu I_q & G_2^TR & G_2^TV \\ * & * & * & -R & 0 \\ * & * & * & * & -V \end{bmatrix},$$

$$\Sigma_2 \triangleq \begin{bmatrix} \lambda R & * & * & * \\ 0 & \lambda V & * & * \\ 0 & 0 & (\gamma-\mu)I_q & * \\ C_1-D_fC_2-T & T & D_{11}-D_fD_2 & \gamma I_m \end{bmatrix}$$

(b) If (10a,b) is satisfied, the matrices of an admissible $L_\infty$-induced deconvolution filter in the form of (2) can be extracted using the following equations:

$$A_f = -V^{-1}S, \quad B_f = -V^{-1}Z, \quad C_f = T \tag{11}$$

Needless to say, $D_f$ is obtained of (10a,b).

**Proof.** See Appendix C.

Due to the fact that LMIs (10a,b) are affine in the system parameters, Theorem 1 can be extended for the case which these parameters are uncertain and lie in the polytope (6).

**Corollary 1.** (*Robust stochastic $L_\infty$-induced deconvolution filtering*) Consider the ESMS system (1a-c) over the polytope (6) and the deconvolution filter of (2). Given $\gamma > 0$ and $\lambda > 0$, The filtering error system (4) is ESMS, and $J_S$ is negative for all nonzero $\omega(t) \in L_{\mathcal{G}_t}^\infty(\mathfrak{R}^q)$ if (10a,b) is satisfied by a single set of $(R, V, Z, S, T, D_f, \mu)$ for all the polytope vertices. In the latter case, the robust deconvolution filter matrices are obtained via (11).

## 5. Improved robust stochastic $L_\infty$-induced deconvolution filter

In this section, an improved robust $L_\infty$-induced deconvolution filter is designed by using Lemma 2 and an idea of structured parameter-dependent matrices. To facilitate the presentation of our results, first consider the stochastic system (1) is without polytopic uncertainties. The following theorem provides sufficient conditions for the existence of admissible peak-to-peak filters for stochastic systems without deterministic uncertainties.

**Theorem 2.** We consider the ESMS system (1a-c) without polytopic uncertainties and the deconvolution filter of (2), and let $\gamma > 0$ and $\lambda > 0$ be given constants. For a sufficiently small scalar $\varepsilon > 0$, the following holds:

(a) The filtering error system (4) is ESMS, and $J_S$ of (3) is negative for all nonzero $\omega(t) \in L_{\mathcal{G}_t}^\infty(\mathfrak{R}^q)$, if there exist $\bar{Q} \triangleq \begin{bmatrix} \bar{Q}_1 & \bar{Q}_2 \\ * & \bar{Q}_3 \end{bmatrix} > 0$, $\bar{R} \in \mathfrak{R}^{n \times n}$, $\bar{S} \in \mathfrak{R}^{n \times n}$, $\bar{T} \in \mathfrak{R}^{n \times n}$, $\bar{A}_f \in \mathfrak{R}^{n \times n}$, $\bar{B}_f \in \mathfrak{R}^{n \times r}$, $\bar{C}_f \in \mathfrak{R}^{m \times n}$, $\bar{D}_f \in \mathfrak{R}^{m \times r}$, and positive scalar $\mu$ so that $\mu < \gamma$, such that the following two LMIs are satisfied:

$$\begin{aligned} \Pi_1(\bar{Q}, \bar{C}_f, \bar{D}_f, \mu) &> 0, \\ \Pi_2(\bar{Q}, \bar{R}, \bar{S}, \bar{T}, \bar{A}_f, \bar{B}_f, \mu) &< 0, \end{aligned} \tag{12a,b}$$

where

$$\Pi_1 \triangleq \begin{bmatrix} \lambda \bar{Q}_1 & \lambda \bar{Q}_2 & 0 & C_1^T - C_2^T \bar{D}_f^T \\ * & \lambda \bar{Q}_3 & 0 & -\bar{C}_f^T \\ * & 0 & (\gamma - \mu) I_q & D_{11}^T - D_2^T \bar{D}_f^T \\ * & * & * & \gamma I_m \end{bmatrix},$$

$$\Pi_2 \triangleq \begin{bmatrix} \Pi_1 & \Pi_2 & \Pi_4 & \Pi_5 & \Pi_8 & 0 & 0 \\ * & \Pi_3 & \Pi_6 & \Pi_7 & \Pi_9 & 0 & 0 \\ * & * & -\bar{Q}_1 & -\bar{Q}_2 & 0 & \sqrt{\varepsilon}G_1^T \bar{R} & \sqrt{\varepsilon}G_1^T \bar{S} \\ * & * & * & -\bar{Q}_3 & 0 & 0 & 0 \\ * & * & * & * & -\mu I_q & G_2^T \bar{R} & G_2^T \bar{S} \\ * & * & * & * & * & \Pi_1 & \Pi_2 \\ * & * & * & * & * & * & \Pi_3 \end{bmatrix},$$

$\Pi_1 = \bar{Q}_1 - \bar{R} - \bar{R}^T$, $\Pi_2 = \bar{Q}_2 - \bar{T} - \bar{S}$,
$\Pi_3 = \bar{Q}_3 - \bar{T} - \bar{T}^T$, $\Pi_4 = (1+\lambda\varepsilon/2)\bar{R}^T + \varepsilon\bar{R}^T A + \varepsilon\bar{B}_f C_2$,
$\Pi_5 = (1+\lambda\varepsilon/2)\bar{T} + \varepsilon\bar{A}_f$, $\Pi_6 = (1+\lambda\varepsilon/2)\bar{S}^T + \varepsilon\bar{S}^T A + \varepsilon\bar{B}_f C_2$,
$\Pi_7 = (1+\lambda\varepsilon/2)\bar{T}^T + \varepsilon\bar{A}_f$, $\Pi_8 = \sqrt{\varepsilon}\bar{R}^T B_1 + \sqrt{\varepsilon}\bar{B}_f D_2$,
$\Pi_9 = \sqrt{\varepsilon}\bar{S}^T B_1 + \sqrt{\varepsilon}\bar{B}_f D_2$

(b) If (12a,b) is satisfied, the matrices of an admissible improved $L_\infty$-induced deconvolution filter in the form of (2) can be extracted using the following equations:

$$\begin{bmatrix} A_f & B_f \\ C_f & D_f \end{bmatrix} = \begin{bmatrix} \bar{T}^{-1} & 0 \\ 0 & I \end{bmatrix} \begin{bmatrix} \bar{A}_f & \bar{B}_f \\ \bar{C}_f & \bar{D}_f \end{bmatrix} \qquad (13)$$

**Proof.** See Appendix D.

**Remark 5.** It is worth noting that for the uncertain case if we solve the robust filter design problem by following the idea in [25], we need to set the general-structured matrix $W$ in Lemma 2 to be constant for the entire uncertainty domain. However, it is observed from the proof of Theorem 2 that when $(A, B_1, C_1, C_2, D_{11}, D_2, G_1, G_2) \in \Omega$ represents an uncertain system, we only need to set part of $W(\alpha)$ to be constant for the entire uncertainty domain. More specifically, for the uncertain case, we select the following structured $W(\alpha)$:

$$W(\alpha) \triangleq \begin{bmatrix} W_1(\alpha) & W_2(\alpha) \\ W_4 & W_3 \end{bmatrix}$$

By following similar lines as in the proof of Theorem 2, we obtain the following corollary.

**Corollary 2.** We consider the ESMS system (1a-c) over the polytope (6) and the deconvolution filter of (2), and let $\gamma > 0$ and $\lambda > 0$ be given constants. For a sufficiently small scalar $\varepsilon > 0$, the following holds:

(a) The filtering error system (4) is ESMS, and $J_S$ of (3) is negative for all nonzero $\omega(t) \in L_\infty^\infty(\mathfrak{R}^q)$, if there exist

$\bar{Q}(\alpha) \triangleq \begin{bmatrix} \bar{Q}_1(\alpha) & \bar{Q}_2(\alpha) \\ * & \bar{Q}_3(\alpha) \end{bmatrix} > 0$, $\bar{R}(\alpha) \in \mathfrak{R}^{n \times n}$, $\bar{S}(\alpha) \in \mathfrak{R}^{n \times n}$, $\bar{T} \in \mathfrak{R}^{n \times n}$, $\bar{A}_f \in \mathfrak{R}^{n \times n}$, $\bar{B}_f \in \mathfrak{R}^{n \times r}$, $\bar{C}_f \in \mathfrak{R}^{m \times n}$, $\bar{D}_f \in \mathfrak{R}^{m \times r}$, and positive scalar $\mu$ so that $\mu < \gamma$, such that the following two LMIs are satisfied:

$$\Theta\left(\bar{Q}(\alpha),\bar{C}_f,\bar{D}_f,\mu\right) > 0,$$
$$\Xi\left(\bar{Q}(\alpha),\bar{R}(\alpha),\bar{S}(\alpha),\bar{T},\bar{A}_f,\bar{B}_f,\mu\right) < 0$$
(14a,b)

where

$$\Theta \triangleq \begin{bmatrix} \lambda\bar{Q}_1(\alpha) & \lambda\bar{Q}_2(\alpha) & 0 & C_1^T - C_2^T\bar{D}_f^T \\ * & \lambda\bar{Q}_3(\alpha) & 0 & -\bar{C}_f^T \\ * & * & (\gamma-\mu)I_q & D_{11}^T - D_2^T\bar{D}_f^T \\ * & * & * & \gamma I_m \end{bmatrix},$$

$$\Xi \triangleq \begin{bmatrix} \bar{\Pi}_1 & \bar{\Pi}_2 & \bar{\Pi}_4 & \bar{\Pi}_5 & \bar{\Pi}_8 & 0 & 0 \\ * & \bar{\Pi}_3 & \bar{\Pi}_6 & \bar{\Pi}_7 & \bar{\Pi}_9 & 0 & 0 \\ * & * & -\bar{Q}_1 & -\bar{Q}_2 & 0 & \sqrt{\varepsilon}G_1^T\bar{R}(\alpha) & \sqrt{\varepsilon}G_1^T\bar{S}(\alpha) \\ * & * & * & -\bar{Q}_3 & 0 & 0 & 0 \\ * & * & * & * & -\mu I_q & G_2^T\bar{R}(\alpha) & G_2^T\bar{S}(\alpha) \\ * & * & * & * & * & \bar{\Pi}_1 & \bar{\Pi}_2 \\ * & * & * & * & * & * & \bar{\Pi}_3 \end{bmatrix},$$

$\bar{\Pi}_1 = \bar{Q}_1(\alpha) - \bar{R}(\alpha) - \bar{R}^T(\alpha),\ \bar{\Pi}_2 = \bar{Q}_2(\alpha) - \bar{T} - \bar{S}(\alpha),$
$\bar{\Pi}_3 = \bar{Q}_3(\alpha) - \bar{T} - \bar{T}^T,\ \bar{\Pi}_4 = (1+\lambda\varepsilon/2I)\bar{R}^T(\alpha) + \varepsilon\bar{R}^T(\alpha)A + \varepsilon\bar{B}_f C_2,$
$\bar{\Pi}_5 = (1+\lambda\varepsilon/2)\bar{T} + \varepsilon\bar{A}_f,\ \bar{\Pi}_6 = (1+\lambda\varepsilon/2)\bar{S}^T(\alpha) + \varepsilon\bar{S}^T(\alpha)A + \varepsilon\bar{B}_f C_2,$
$\bar{\Pi}_7 = (1+\lambda\varepsilon/2)\bar{T}^T + \varepsilon\bar{A}_f,\ \bar{\Pi}_8 = \sqrt{\varepsilon}\bar{R}^T(\alpha)B_1 + \sqrt{\varepsilon}\bar{B}_f D_2,$
$\bar{\Pi}_9 = \sqrt{\varepsilon}\bar{S}^T(\alpha)B_1 + \sqrt{\varepsilon}\bar{B}_f D_2$

(b) If (14a,b) is satisfied, the matrices of an admissible robust improved $L_\infty$-induced deconvolution filter in the form of (2) are given by (13).

**Remark 6.** The LMI conditions in Corollary 2 still cannot be implemented due to it infinite-dimensional nature in the parameter $\alpha$. Our purpose hereafter is to transform the infinite-dimensional condition in Corollary 2 into finite-dimensional condition that depends only on the vertex matrices of the polytope $\Omega$. Then, we have the main filtering result in the following theorem.

**Theorem 3.** (*Improved robust stochastic $L_\infty$-induced deconvolution filtering*) We consider the ESMS system (1a-c) over the polytope (6) and the deconvolution filter of (2), and let $\gamma > 0$ and $\lambda > 0$ be given constants. For a sufficiently small scalar $\varepsilon > 0$, the following holds:

(a) The filtering error system (4) is ESMS, and $J_S$ of (3) is negative for all nonzero $\omega(t) \in L_{\vartheta_1}^\infty(\mathfrak{R}^q)$, if there exist $\bar{Q}_i \triangleq \begin{bmatrix} \bar{Q}_{1i} & \bar{Q}_{2i} \\ * & \bar{Q}_{3i} \end{bmatrix} > 0$, $\bar{R}_i \in \mathfrak{R}^{n\times n}$, $\bar{S}_i \in \mathfrak{R}^{n\times n}$, $\bar{T} \in \mathfrak{R}^{n\times n}$, $\bar{A}_f \in \mathfrak{R}^{n\times n}$, $\bar{B}_f \in \mathfrak{R}^{n\times r}$, $\bar{C}_f \in \mathfrak{R}^{m\times n}$, $\bar{D}_f \in \mathfrak{R}^{m\times r}$, and positive scalar $\mu$ so that $\mu < \gamma$, satisfying

$$\Theta_i > 0, \quad i = 1,\ldots,s,$$
$$\Xi_{ii} < 0, \quad i = 1,\ldots,s, \qquad (15\text{a-c})$$
$$\Xi_{ij} + \Xi_{ji} < 0, \quad 1 \leq i < j \leq s,$$

where

$$\Theta_i \triangleq \begin{bmatrix} \lambda \bar{Q}_{1i} & \lambda \bar{Q}_{2i} & 0 & C_{1i}^T - C_{2i}^T \bar{D}_f^T \\ * & \lambda \bar{Q}_{3i} & 0 & -\bar{C}_f^T \\ * & 0 & (\gamma - \mu)I_q & D_{11,i}^T - D_{2i}^T \bar{D}_f^T \\ * & * & * & \gamma I_m \end{bmatrix}, \qquad (16)$$

$$E_{ij} \triangleq \begin{bmatrix} \Lambda_1 & \Lambda_2 & \Lambda_4 & \Lambda_5 & \Lambda_8 & 0 & 0 \\ * & \Lambda_3 & \Lambda_6 & \Lambda_7 & \Lambda_9 & 0 & 0 \\ * & * & -\bar{Q}_{1i} & -\bar{Q}_{2i} & 0 & \sqrt{\varepsilon} G_{1j}^T \bar{R}_i & \sqrt{\varepsilon} G_{1j}^T \bar{S}_i \\ * & * & * & -\bar{Q}_{3i} & 0 & 0 & 0 \\ * & * & * & * & -\mu I_q & G_{2j}^T \bar{R}_i & G_{2j}^T \bar{S}_i \\ * & * & * & * & * & \Lambda_1 & \Lambda_2 \\ * & * & * & * & * & * & \Lambda_3 \end{bmatrix}, \qquad (17)$$

$$\begin{aligned}
&\Lambda_1 = \bar{Q}_{1i} - R_i - R_i^T, \quad \Lambda_2 = \bar{Q}_{2i} - \bar{T} - \bar{S}_i, \\
&\Lambda_3 = \bar{Q}_{3i} - \bar{T} - \bar{T}^T, \quad \Lambda_4 = (1 + \lambda \varepsilon / 2)\bar{R}_i^T + \varepsilon \bar{R}_i^T A_j + \varepsilon \bar{B}_f C_{2j}, \\
&\Lambda_5 = (1 + \lambda \varepsilon / 2)\bar{T} + \varepsilon \bar{A}_f, \quad \Lambda_6 = (1 + \lambda \varepsilon / 2)\bar{S}_i^T + \varepsilon \bar{S}_i^T A_j + \varepsilon \bar{B}_f C_{2j}, \\
&\Lambda_7 = (1 + \lambda \varepsilon / 2)\bar{T}^T + \varepsilon \bar{A}_f, \quad \Lambda_8 = \sqrt{\varepsilon}\bar{R}_i^T B_{1j} + \sqrt{\varepsilon}\bar{B}_f D_{2j}, \\
&\Lambda_9 = \sqrt{\varepsilon}\bar{S}_i^T B_{1j} + \sqrt{\varepsilon}\bar{B}_f D_{2j}
\end{aligned} \qquad (18)$$

(b) If (15a-c) is satisfied, the matrices of an admissible robust improved $L_\infty$-induced deconvolution filter in the form of (2) are given by (13).

**Proof.** See Appendix E.

**Remark 7.** Observe that for given $\lambda$ and $\varepsilon$, the conditions in Theorem 3 are LMIs not only over the matrix variables, but also over the scalar $\gamma$. This implies that the scalar $\gamma$ can be included as an optimization variable to obtain a reduction of the upper bound of the peak-to-peak norm of the deconvolution error system (4). Then the minimum (in terms of the feasibility of Theorem 3) upper bound $\gamma$ can be readily found by solving the following convex optimization problem:

Minimize $\gamma$ subject to (15a-c) for sufficiently small $\varepsilon > 0$ and $0 < \lambda < -2\max\left(\text{real}\left\{\text{eig}(\tilde{A})\right\}\right)$

**Remark 8.** The minimization of the upper bound $\gamma$ to the peak-to-peak norm depends on the choice of parameters $\varepsilon$ and $\lambda$ (where $0 < \lambda < -2\max\left(\text{real}\left\{\text{eig}(\tilde{A})\right\}\right)$). The question arises how to find the optimal combination of these parameters in order to obtain a tighter upper bound. The proposed way in [32] to address the tuning issue is to choose for a cost function the parameter $t_{min}$ that is obtained while solving the feasibility problem using Matlab's LMI toolbox [33]. This scalar parameter is positive in cases where the combination of the tuning parameters is one that does not allow a feasible solution to the set of LMIs considered. Applying a numerical optimization algorithm, such as the function *fminsearchbnd* of Matlab [34], to the above cost function, a locally convergent solution to the problem is obtained. If the resulting minimum value of the cost function is negative, the tuning parameters that solve the problem are found.

**Remark 9.** Theorem 1 and Corollary 1 apply only the tuning parameter $\lambda$. The best possible upper bound $\gamma$ is found by combining the minimization of $\gamma(\lambda)$ for fixed $\lambda$ with a line search over $0 < \lambda < -2\max(real\{eig(\tilde{A})\})$ [17,19].

**6. Application to fault reconstruction**

The main contribution of this section is the generalization of the obtained results for sensor fault reconstruction problem in [29] to continuous time stochastic systems by the $L_\infty$-induced deconvolution filter. It is worth mentioning that fault reconstruction is different from the common fault detection and isolation (FDI) methods based on the residual generation techniques in the sense that it not only detects and isolates the fault, but also provides an estimate of the fault. This approach is very useful for incipient (slowly varying) faults, which are very difficult to detect. A detailed description of robust reconstruction and discussion of its importance can be found in [29,35].

With injection a sensor fault in the measured output of the stochastic system (1), we have

$$\begin{aligned} dx(t) &= [Ax(t) + B_1\omega(t)]dt + [G_1x(t) + G_2\omega(t)]d\beta(t), \\ y(t) &= C_2x(t) + D_2\omega(t) + Ff(t) \end{aligned} \quad (19a,b)$$

where $f \in \Re^p$ are the sensor faults. The vector $\omega(t)$ here is assumed to be the exogenous disturbance which presents nonlinearities, unmodeled dynamics and uncertainties [29]. The dimensions of the state, output and fault vectors satisfy $n \geq r > p$. Without loss of generality, it can be assumed that the outputs of the system have been reordered (and scaled if necessary) so that the matrix $F$ has the following structure

$$F = \begin{bmatrix} 0 \\ F_2 \end{bmatrix} \quad (20)$$

where $F_2 \in \Re^{p \times p}$ is nonsingular matrix. This implies that some sensors are not potentially faulty, see Remark 1 in [35]. Scaling the output $y$ and then partitioning appropriately would yield

$$\begin{aligned} y_1 &= C_{21}x + D_{21}\omega, \\ y_2 &= C_{22}x + D_{22}\omega + F_2f \end{aligned} \quad (21a,b)$$

where $y_1 \in \Re^{r-p}$. Notice now that systems (19a) and (21a) make up a fault-free system.

Motivated by [29], to reconstruct the fault $f(t)$ we design a fault estimation filter based on a deconvolution filter for the fault-free system defined by (19a) and (21a), which is of the following form:

$$d\hat{x} = A_f \hat{x} dt + B_f y_1 dt, \quad \hat{x}(0) = 0,$$
$$\hat{y} = C_f \hat{x} + D_f y_1, \quad (22)$$
$$\hat{f} = H(y - \hat{y}),$$

where $\hat{x} \in \Re^n$, $\hat{y} \in \Re^r$, and $\hat{f} \in \Re^p$ are the state, output, and fault estimation vectors of the fault estimation filter. The matrix $H$ is defined as $H = \begin{bmatrix} H_1 & F_2^{-1} \end{bmatrix}$ with $H_1 \in \Re^{p \times (r-p)}$ being a weighting matrix.

Defining $e_f \triangleq f - \hat{f}$ as the error in fault reconstruction, denoting $\xi = \begin{bmatrix} x^T & \hat{x}^T \end{bmatrix}^T$, and combining (19a,b), (21a), and (22), we have the following augmented system:

$$d\xi = \tilde{A} \xi dt + \tilde{B} \omega dt + \left[ \tilde{G}_1 \xi + \tilde{G}_2 \omega \right] d\beta,$$
$$e_f = \tilde{C} \xi + \tilde{D} \omega \quad (23)$$

where

$$\tilde{A} = \begin{bmatrix} A & 0 \\ B_f C_{21} & A_f \end{bmatrix}, \quad \tilde{B} = \begin{bmatrix} B_1 \\ B_f D_{21} \end{bmatrix}, \quad \tilde{G}_1 = \begin{bmatrix} G_1 & 0 \\ 0 & 0 \end{bmatrix},$$
$$\tilde{G}_2 = \begin{bmatrix} G_2 \\ 0 \end{bmatrix}, \quad \tilde{C} = \begin{bmatrix} H(D_f C_{21} - C_2) & HC_f \end{bmatrix}, \quad \tilde{D} = H(D_f D_{21} - D_2) \quad (24)$$

The objective now would be to minimize the effect of $\omega$ on the reconstruction error $e_f$. This can be achieved using the Lemma 2. Defining the following performance index

$$J_F \triangleq \|e_f\|_\infty - \gamma \|\omega\|_\infty \quad (25)$$

Our objective is finding a general type filter of the form (22) that leads to an ESMS reconstruction error $e_f$ such that $J_F$ of (25) is negative for all nonzero $\omega(t) \in L_{\mathcal{F}_t}^\infty(\Re^q)$.

**Theorem 4.** We consider the ESMS system (19a,b) and the fault estimation filter of (22), and let $\gamma > 0$ and $\lambda > 0$ be given constants. For a sufficiently small scalar $\varepsilon > 0$, the following holds:

(a) The reconstruction error system (23) is ESMS, and $J_F$ of (25) is negative for all nonzero $\omega(t) \in L_{\mathcal{F}_t}^\infty(\Re^q)$, if there exist $\bar{Q} \triangleq \begin{bmatrix} \bar{Q}_1 & \bar{Q}_2 \\ * & \bar{Q}_3 \end{bmatrix} > 0$, $\bar{R} \in \Re^{n \times n}$, $\bar{S} \in \Re^{n \times n}$, $\bar{T} \in \Re^{n \times n}$, $\bar{A}_f \in \Re^{n \times n}$, $\bar{B}_f \in \Re^{n \times (r-p)}$, $\bar{C}_f \in \Re^{r \times n}$, $\bar{D}_f \in \Re^{r \times (r-p)}$, and positive scalar $\mu$ so that $\mu < \gamma$, satisfying

$$\Pi(\bar{Q}, \bar{C}_f, \bar{D}_f, \mu) > 0,$$
$$\mho(\bar{Q}, \bar{R}, \bar{S}, \bar{T}, \bar{A}_f, \bar{B}_f, \mu) < 0, \quad (26a,b)$$

where

$$\Pi \triangleq \begin{bmatrix} \lambda \bar{Q}_1 & * & * & * \\ \lambda \bar{Q}_2^T & \lambda \bar{Q}_3 & * & * \\ 0 & 0 & (\gamma - \mu) I_q & * \\ H(\bar{D}_f C_{21} - C_2) & H\bar{C}_f & H(\bar{D}_f D_{21} - D_2) & \gamma I_p \end{bmatrix},$$

$$\mho \triangleq \begin{bmatrix} \Phi_1 & \Phi_2 & \Phi_4 & \Phi_5 & \Phi_8 & 0 & 0 \\ * & \Phi_3 & \Phi_6 & \Phi_7 & \Phi_9 & 0 & 0 \\ * & * & -\bar{Q}_1 & -\bar{Q}_2 & 0 & \sqrt{\varepsilon}G_1^T \bar{R} & \sqrt{\varepsilon}G_1^T \bar{S} \\ * & * & * & -\bar{Q}_3 & 0 & 0 & 0 \\ * & * & * & * & -\mu I_q & G_2^T \bar{R} & G_2^T \bar{S} \\ * & * & * & * & * & \Phi_1 & \Phi_2 \\ * & * & * & * & * & * & \Phi_3 \end{bmatrix},$$

$\Phi_1 = \bar{Q}_1 - R - R^T$, $\Phi_2 = \bar{Q}_2 - \bar{T} - \bar{S}$,
$\Phi_3 = \bar{Q}_3 - \bar{T} - \bar{T}^T$, $\Phi_4 = (1+\lambda\varepsilon/2)R^T + \varepsilon R^T A + \varepsilon \bar{B}_f C_{21}$,
$\Phi_5 = (1+\lambda\varepsilon/2)\bar{T} + \varepsilon \bar{A}_f$, $\Phi_6 = (1+\lambda\varepsilon/2)\bar{S}^T + \varepsilon \bar{S}^T A + \varepsilon \bar{B}_f C_{21}$,
$\Phi_7 = (1+\lambda\varepsilon/2)\bar{T}^T + \varepsilon \bar{A}_f$, $\Phi_8 = \sqrt{\varepsilon}R^T B_1 + \sqrt{\varepsilon}\bar{B}_f D_{21}$,
$\Phi_9 = \sqrt{\varepsilon}\bar{S}^T B_1 + \sqrt{\varepsilon}\bar{B}_f D_{21}$

(b) If (26a,b) is satisfied, the matrices of an admissible robust fault estimation filter in the form of (22) are given by (13).

The proof of Theorem 4 can be established based on Lemma 2 and using similar techniques to those in Appendix D, and thus is omitted here

## 7. Illustrative examples

In this section, we provide two examples to demonstrate the effectiveness of the proposed design methods. Example 1 is for the unknown input estimation problem, while Example 2 is for the fault estimation problem.

### 7.1. Unknown input estimation

In this subsection, we use the following example, which is taken from [10], to demonstrate the usefulness of two proposed approaches for the deconvolution problem. Consider the linear continuous-time stochastic system (1a-c) with the parameters as follows:

$$A = \begin{bmatrix} -0.1 & 3+0.5a \\ -3 & -4 \end{bmatrix}, \quad B_1 = \begin{bmatrix} -0.5a \\ 0.9a \end{bmatrix}, \quad G_1 = \begin{bmatrix} 0.5 & 0 \\ 0 & 0.5 \end{bmatrix},$$
$$G_2 = \begin{bmatrix} 0.1 \\ 0.1 \end{bmatrix}, \quad C_2 = [0.8 \quad 0.8(1+a)], \quad D_2 = 0.45 - 0.5a, \quad (27)$$
$$C_1 = [0 \quad 0], \quad D_{11} = 1$$

where $a$ is a bounded constant uncertain parameter satisfying $|a| \leq 0.3$. This uncertain system can be modeled with a two-vertex polytope.

We solve the filtering problem for this system by two approaches described as follows:

1. (*Quadratic stability approach*) By Corollary 1, the obtained minimum upper bound to the $L_\infty$-induced norm of the system (4) with the line search on $\lambda$ is $\gamma = 0.7278$ for ($\lambda = 2.5$), and also we obtain $\mu = 0.4113$ and the associated matrices for the deconvolution filter as follows:

$$A_f = \begin{bmatrix} -2.7521 & 0.2916 \\ -4.8316 & -3.8401 \end{bmatrix}, \quad B_f = \begin{bmatrix} 0.6383 \\ 0.4485 \end{bmatrix} \qquad (28)$$
$$C_f = [-0.2142 \quad -0.2326], \quad D_f = 2.3112$$

2. (*Structured parameter-dependent approach*) By Theorem 3, the obtained minimum upper bound of the peak-to-peak gain is $\gamma = 0.6932$ for ($\lambda = 2.7$ and $\varepsilon = 0.001$), and also we obtain $\mu = 0.3793$ and the associated matrices for the deconvolution filter as follows:

$$A_f = \begin{bmatrix} -0.0087 & 2.7929 \\ -3.3489 & -3.8237 \end{bmatrix}, \quad B_f = \begin{bmatrix} 0.1717 \\ -0.4366 \end{bmatrix} \qquad (29)$$
$$C_f = [0.4156 \quad 0.4031], \quad D_f = 2.3310$$

The above calculated results show that for this example, the robust filtering result in the quadratic framework is conservative than the structured parameter-dependent approach. Note that, how it was written in Remark 8, a better selection of the scalar $\lambda$ and $\varepsilon$ can be done using the Matlab function *fminsearchbnd*.

*7.2. Fault estimation*

In this subsection, the stochastic peak-to-peak fault estimation is exploited for reconstruction the sensor fault in an inverted pendulum example which is taken from [36]. We consider a single degree of freedom inverted pendulum system with the multiplicative white noise, which has been stabilized with a pre-designed controller using the stochastic nonlinear $H_\infty$ state feedback method [36]. The model of the inverted pendulum system is given by

$$ml^2 \ddot{\theta} - mgl \sin\theta + (\varsigma + \eta)\dot{\theta} + \kappa\theta = u + \omega_1 \qquad (30)$$

where $\kappa$ is the spring coefficient, $\varsigma$ is damping coefficient, and $\eta$ is the stochastic uncertainty in the damping. $\theta$ is the inclination angle of the pendulum, $l$ and $m$ are its length and mass, respectively, and g is the gravitational acceleration. $u$ is the control input that has been pre-designed as $u = k_1\theta + k_2 ml^2\dot{\theta}$, and $\omega_1$ is the exogenous disturbance acting on the control input $u$, which is considered as a zero-mean white noise with intensity $R_1$.

A linearization of the nonlinear system (30) has been made about the equilibrium point at the origin. The state variables $x_1$ and $x_2$ are defined as $x_1 = \theta$ and $x_2 = ml^2\dot{\theta}$. The incipient fault $f$ is injected to the angle $\theta$. The measured outputs are $y_1 = x_2 + \omega_2$ and $y_2 = x_1 + f$, where $\omega_2$ is a zero-mean white noise with intensity $R_2$. Supposing that the exogenous disturbance is $\omega = [\omega_1 \quad \omega_2]^T$, and $d\beta(t) = \eta(t)dt$ that the Wiener process $\beta(t)$ has defined in Section 2, the following state space matrices of the system are obtained regarding the notation associated with (19a,b):

$$A = \begin{bmatrix} 0 & \frac{1}{ml^2} \\ -\kappa + mgl + k_1 & -\frac{\varsigma}{ml^2} + k_2 \end{bmatrix}, \quad B_1 = \begin{bmatrix} 0 & 0 \\ \sqrt{R_1} & 0 \end{bmatrix},$$

$$G_1 = \begin{bmatrix} 0 & 0 \\ 0 & -\frac{1}{ml^2} \end{bmatrix}, \quad G_2 = B_1 \times 0, \quad C_2 = \begin{bmatrix} 0 & 1 \\ 1 & 0 \end{bmatrix},$$

$$D_2 = \begin{bmatrix} 0 & \sqrt{R_2} \\ 0 & 0 \end{bmatrix}, \quad F = \begin{bmatrix} 0 & 1 \end{bmatrix}^T$$

The system parameters are $m = 0.5$ kg, $l = 0.7$ m, $\kappa = 0.5$ N/m, and $\varsigma = 0.25$. The intensities $R_1$ and $R_2$ are 1 and 0.4, respectively. The controller parameters are $k_1 = -29.7398$ and $k_2 = -63.9668$. By Theorem 4, with choosing $H_1 = 1$, $\gamma = 1$, $\lambda = 2$, and $\varepsilon = 0.001$, we obtain $\mu = 0.8453$ and the fault estimation filter matrices as follows:

$$A_f = \begin{bmatrix} -0.6364 & 2.8163 \\ -23.3901 & -79.5138 \end{bmatrix}, \quad B_f = \begin{bmatrix} 0.0056 \\ -0.0650 \end{bmatrix}, \quad\quad (31)$$
$$C_f = \begin{bmatrix} 0.1048 & -0.0344 \\ 0.2515 & 0.0294 \end{bmatrix}, \quad D_f = \begin{bmatrix} 0.5073 \\ 0.4925 \end{bmatrix}$$

The fault estimation filter (22) is used for reconstruction of the sensor fault $f(t)$ in the nonlinear stochastic system (30). The nonlinear simulation result is shown in Fig. 1. From Fig. 1, it is obvious that the proposed fault reconstruction scheme reconstructs the fault perfectly when sensor fault is applied in the presence of the multiplicative noise $\eta$ and the additive noises $\omega_1$ and $\omega_2$.

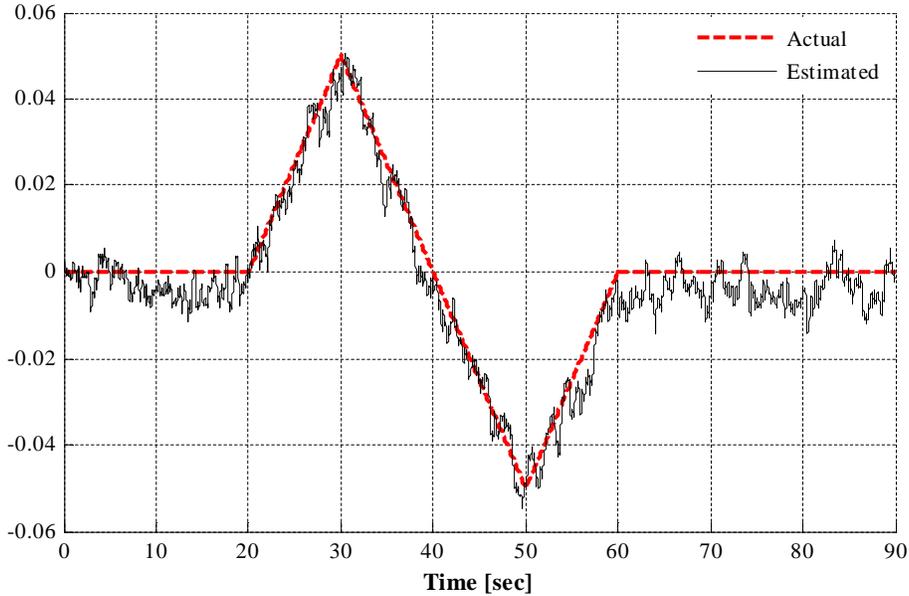

**Fig. 1** Sensor fault reconstruction on the angle sensor of inverted pendulum nonlinear model.

## 8. Conclusions

In the present paper, we have discussed the problem of peak-to-peak deconvolution filtering for stationary uncertain continuous-time linear stochastic systems with state- and input-dependent noise. Based on proposed $L_\infty$-induced performance characterizations, sufficient conditions for the existence of two desired deconvolution filters are presented in terms of LMIs, which guarantee the mean-square exponential stability of filtering error dynamic and satisfy the given peak-to-peak performance index. The obtained results from the parameter-dependent stability based approach are less conservative than the existing one in the quadratic framework. The improved robust $L_\infty$-induced filtering technique is used to design a fault estimation filter. Two simulation examples have been exploited to demonstrate the effectiveness of the proposed design procedures.

## Appendix A. Proof of Lemma 1

**Proof.** This Lemma can be proven through a trend similar to the proof of Lemma 1 in [22].

First, we can follow from (A.3) that $\tilde{A}^T Q + Q\tilde{A} + \tilde{G}_1^T Q \tilde{G}_1 < 0$. Therefore, we conclude from Proposition 1 that the system (4) is ESMS. We realize (4) in the following form:

$$d\xi(t) = a(x,t)dt + \sigma(x,t)d\tilde{\beta}(t) \tag{A.1}$$

where $a(x,t) = \tilde{A}x + \tilde{B}\omega$, $\sigma(x,t) = [\tilde{G}_1 x \quad \tilde{G}_2 \omega]$, and $d\tilde{\beta} = [d\beta \quad d\beta]^T$. $\tilde{\beta}(t)$ is the augmented Wiener process in $\Re^2$.

We consider the positive definite function $V(\xi) = \xi^T Q \xi$ where $Q > 0$ is a constant matrix. Applying the *Itô* formula (see [30]) to evaluate differential of the quadratic form $\xi^T Q \xi$ and taking expectation, we obtain

$$\begin{aligned} E\{dV(\xi)\} &= E\left\{V_\xi a dt + \frac{1}{2} Tr\left[\sigma^T V_{\xi\xi} \sigma \tilde{Q}\right] dt\right\} \\ &= E\left\{\xi^T Q(\tilde{A}\xi + \tilde{B}\omega) + (\tilde{A}\xi + \tilde{B}\omega)^T Qx + \xi^T \tilde{G}_1^T Q \tilde{G}_1 \xi + \omega^T \tilde{G}_2^T Q \tilde{G}_1 \xi + x^T \tilde{G}_1^T Q \tilde{G}_2 \xi + \omega^T \tilde{G}_2^T Q \tilde{G}_2 \omega\right\} dt \end{aligned} \tag{A.2}$$

where $\tilde{Q} \triangleq I_2$ is the covariance matrix of $\tilde{\beta}(t)$.

The inequality (7) can be rewritten using Schur complements as follows:

$$\Gamma \triangleq \begin{bmatrix} \tilde{A}^T Q + Q\tilde{A} + \lambda Q + \tilde{G}_1^T Q \tilde{G}_1 & Q\tilde{B} + \tilde{G}_1^T Q \tilde{G}_2 \\ \tilde{B}^T Q + \tilde{G}_2^T Q \tilde{G}_1 & -\mu I_q + \tilde{G}_2^T Q \tilde{G}_2 \end{bmatrix} < 0 \tag{A.3}$$

Defining $\kappa \triangleq [\xi^T, \omega^T]^T$ and denoting $Z(t) = E\{V(\xi(t))\}$ we obtain from (A.3) that $E\{\kappa^T \Gamma \kappa\} \leq 0$ which implies by (A.2) that

$$\dot{Z}(t) + \lambda Z(t) - \mu E\{\|\omega(t)\|^2\} \leq 0, \quad \forall t \geq 0, \quad Z(0) = 0 \tag{A.4}$$

Multiplying (A.4) by $e^{\lambda t}$, we get

$$\frac{d}{dt}\left[e^{\lambda t}Z(t)\right] \leq \mu e^{\lambda t}E\left\{\|\omega(t)\|^2\right\} \Rightarrow e^{\lambda t}Z(t) \leq \mu\int_0^t e^{\lambda\tau}E\left\{\|\omega(\tau)\|^2\right\}d\tau \Rightarrow Z(t) \leq \mu\int_0^t e^{-\lambda(t-\tau)}E\left\{\|\omega(t)\|^2\right\}dt$$
$$\Rightarrow Z(t) \leq \mu\sup_{t\geq 0}\left[E\left\{\|\omega(t)\|^2\right\}\right]\int_0^t e^{-\lambda(t-\tau)}dt \Rightarrow Z(t) \leq \frac{\mu}{\lambda}\|\omega\|_\infty^2(1-e^{-\lambda t})$$

(A.5)

Ultimately we obtain that

$$Z(t) \leq \frac{\mu}{\lambda}\|\omega\|_\infty^2 \tag{A.6}$$

On the other hand, applying the Schur complement we have from (8)

$$\kappa^T\left(\begin{bmatrix}\lambda Q & 0 \\ 0 & (\gamma-\mu)I\end{bmatrix} - \gamma^{-1}\begin{bmatrix}\tilde{C}^T \\ \tilde{D}^T\end{bmatrix}\begin{bmatrix}\tilde{C} & \tilde{D}\end{bmatrix}\right)\kappa > 0 \tag{A.7}$$

Taking the expectation of the latter inequality, we find that

$$\lambda Z(t) + (\gamma-\mu)E\left\{\|\omega(t)\|^2\right\} > \gamma^{-1}E\left\{\|z(t)\|^2\right\}, \quad \forall t \geq 0 \tag{A.8}$$

Using (A.6) in (A.8) and taking the supremum over time $t \geq 0$ yields $\|z\|_\infty < \gamma\|\omega\|_\infty$.

Finally, the first diagonal block in (7) implies that $\tilde{A}^T Q + Q\tilde{A} + \lambda Q < 0$. It is readily obtained $\tilde{A} + \frac{1}{2}\lambda I < 0$, i.e.,

$\begin{bmatrix} A + 1/2\lambda I & 0 \\ B_f C_2 & A_f + 1/2\lambda I \end{bmatrix} < 0$. Hence $\lambda$ should necessarily be chosen in such a way that $\lambda \in \left(0, -2\max\left(real\left\{eig(\tilde{A})\right\}\right)\right)$, i.e.,

$\lambda \in \left(0, -2\max\left(real\left\{eig(A), eig(A_f)\right\}\right)\right)$.

**Appendix B. Proof of Lemma 2**

**Proof.** This Lemma can be proven through a trend similar to the proof of Proposition 3.1 in [5].

The Lemma is proved by showing the equivalence between (7) and (9). First, inequality (9) is equivalent to

$$\begin{bmatrix} -Q & Q(I+\varepsilon\hat{A}) & \sqrt{\varepsilon}Q\tilde{B} & 0 \\ * & -Q & 0 & \sqrt{\varepsilon}\tilde{G}_1^T Q \\ * & * & -\mu I & \tilde{G}_2^T Q \\ * & * & * & -Q \end{bmatrix} < 0 \tag{B.1}$$

where $\hat{A} = \tilde{A} + \frac{\lambda}{2}I$. The equivalence between (B.1) and (9) can be proved as follows. On one hand, if LMI (B.1) holds, LMI (7) is readily established by choosing $W = W^T = Q$. On the other hand, if LMI (7) holds, we can explore the fact that $W + W^T - Q > 0$ and $Q > 0$, so that $W$ is nonsingular. In addition, we have $(Q - W^T)Q^{-1}(Q - W) \geq 0$, which implies $-W^T Q^{-1} W \leq Q - W - W^T$. Therefore, we can conclude from (9) that

$$\begin{bmatrix} -W^TQ^{-1}W & W^T(I+\varepsilon\hat{A}) & \sqrt{\varepsilon}W^T\tilde{B} & \sqrt{\varepsilon}\tilde{G}_1^TW \\ * & -Q & 0 & \tilde{G}_2^TW \\ * & * & -\mu I & 0 \\ * & * & * & -W^TQ^{-1}W \end{bmatrix} \qquad (B.2)$$

Performing a congruence transformation to (B.2) by $diag(W^{-1}Q, I, I, W^{-1}Q)$ yields (B.1).

Now, performing a congruence transformation to (B.1) by $diag(I, \varepsilon^{-1/2}I, I, I)$, we obtain

$$\begin{bmatrix} -Q & Q(\varepsilon^{-1/2}I + \varepsilon^{1/2}\hat{A}) & \sqrt{\varepsilon}Q\tilde{B} & 0 \\ * & -\varepsilon^{-1}Q & 0 & \tilde{G}_1^TQ \\ * & * & -\mu I & \tilde{G}_2^TQ \\ * & * & * & -Q \end{bmatrix} < 0 \qquad (B.3)$$

by Schur complements formula, (B.3) is equivalent to

$$\begin{bmatrix} Q\hat{A} + \hat{A}^TQ + \varepsilon\hat{A}^TQ\hat{A} & Q\tilde{B} + \varepsilon\hat{A}^TQ\tilde{B} & \tilde{G}_1^TQ \\ * & -\mu I + \varepsilon\tilde{B}^TQ\tilde{B} & \tilde{G}_2^TQ \\ * & * & -Q \end{bmatrix} < o \qquad (B.4)$$

which is further equivalent to

$$\begin{bmatrix} \tilde{A}^TQ + Q\tilde{A} + \lambda Q & Q\tilde{B} & \tilde{G}_1^TQ \\ \tilde{B}^TQ & -\mu I_q & \tilde{G}_2^TQ \\ Q\tilde{G}_1 & Q\tilde{G}_2 & -Q \end{bmatrix} + \varepsilon \begin{bmatrix} \hat{A}^T \\ \hat{B}^T \\ 0 \end{bmatrix} Q \begin{bmatrix} \hat{A} & \hat{B} & 0 \end{bmatrix} < 0 \qquad (B.5)$$

Since $Q > 0$ and $\varepsilon$ is sufficiently small positive, (B.5) is in fact equivalent to (7), and the proof is completed.

**Appendix C. Proof of Theorem 1**

**Proof.** This theorem can be proved through a trend similar to those used in [1,22].

(a) According to Lemma 1, the filtering error system (4) is ESMS, and $J_s$ of (3) is negative for all nonzero $\omega(t) \in L_{\mathfrak{R}_1}^{\infty}(\mathfrak{R}^q)$ if there exist $Q > 0$, $\lambda \in (0, -2\max(real\{eig(\tilde{A})\}))$, and $\mu > 0$ so that $\mu < \gamma$, satisfying (7) and (8). Let $Q$ and $Q^{-1}$ be partitioned in the form of $Q \triangleq \begin{bmatrix} X & M \\ M^T & U \end{bmatrix}$ and $Q^{-1} \triangleq \begin{bmatrix} Y & N \\ N^T & V \end{bmatrix}$, where we require that $X > Y^{-1}$. Define matrices $J \triangleq \begin{bmatrix} Y & I_n \\ N^T & 0 \end{bmatrix}$, $\tilde{J} \triangleq diag[J, I_q, J]$, and $\bar{J} \triangleq diag[J, I_q, I_m]$. Performing congruence transformations to (7) by $\tilde{J}$ and to (8) by $\bar{J}$, respectively, and taking into account (5), and carrying out some multiplications and through the substitution of:

$$Z \triangleq MB_f, \quad \tilde{Z} \triangleq C_f N^T, \quad \hat{Z} \triangleq MA_f N^T \qquad (C.1)$$

(C.2) and (C.3) are obtained

$$\begin{bmatrix} AY+YA^T+\lambda Y & * & * & * & * \\ A^T+XAY+ZC_2Y+\lambda I+\hat{Z} & XA+A^TX+\lambda X+C_2^TZ^T+ZC_2 & * & * & * \\ B_1^T & B_1^TX+D_2^TZ^T & -\mu I_q & * & * \\ G_1Y & G_1 & G_2 & -Y & * \\ XG_1Y & XG_1 & XG_2 & -I & -X \end{bmatrix} < 0 \qquad (C.2)$$

$$\begin{bmatrix} \lambda Y & * & * & * \\ \lambda I & \lambda X & * & * \\ 0 & 0 & (\gamma-\mu)I_q & * \\ C_1Y-D_fC_2Y-\tilde{Z} & C_1-D_fC_2 & D_{11}-D_fD_2 & \gamma I_m \end{bmatrix} > 0 \qquad (C.3)$$

Defining $\Upsilon$ and $\bar{\Upsilon}$ as below,

$$\Upsilon \triangleq diag\left\{\begin{bmatrix} R & 0 \\ -R & I_n \end{bmatrix}, I_q, \begin{bmatrix} R & 0 \\ -R & I_n \end{bmatrix}\right\},$$

$$\bar{\Upsilon} \triangleq diag\left\{\begin{bmatrix} R & 0 \\ -R & I_n \end{bmatrix}, I_q, I_m\right\}, \qquad (C.4)$$

and substituting

$$S \triangleq \hat{Z}R, \quad T \triangleq \tilde{Z}R, \quad R \triangleq Y^{-1}, \quad V = X - R \qquad (C.5)$$

and performing congruence transformations to (C.2) by $\Upsilon^T$ and to (C.3) by $\bar{\Upsilon}^T$, respectively, (10a) and (10b) are achieved.

(b) If there exists a solution to (10a,b), from (C.1) we obtain that

$$A_f = M^{-1}\hat{Z}N^{-T}, \quad B_f = M^{-1}Z, \quad C_f = \tilde{Z}N^{-T} \qquad (C.6)$$

Applying (C.6) in the transfer function matrix of the deconvolution filter, which is obtained of (2), we find that

$$\begin{aligned} H_{\hat{z}y}(s) &= C_f(sI-A_f)^{-1}B_f + D_f \\ &= \tilde{Z}(sMN^T-\hat{Z})^{-1}Z + D_f \\ &= \tilde{Z}(s(I_n-XY)-\hat{Z})^{-1}Z + D_f \end{aligned} \qquad (C.7)$$

Now regarding (C.5), $H_{\hat{z}y}(s)$ is obtained as the following:

$$\begin{aligned} H_{\hat{z}y}(s) &= T(s(R-X)-S)^{-1}Z + D_f \\ &= T(sI-(R-X)^{-1}S)^{-1}(R-X)^{-1}Z + D_f \end{aligned} \qquad (C.8)$$

which means (11) is established and the proof is completed.

**Appendix D. Proof of Theorem 2**

**Proof.** This theorem can be proved by following similar lines as in the proof of Proposition 3.2 in [5].

(a) According to Lemma 2, the filtering error system (4) is ESMS, and $J_S$ of (3) is negative for all nonzero $\omega(t) \in L^\infty_{\mathcal{S}_t}(\mathfrak{R}^q)$

if there exist $Q > 0$, $\lambda \in (0, -2\max(real\{eig(A)\}))$, and $\mu > 0$ so that $\mu < \gamma$, satisfying (8) and (9). Let $Q$ and $W$ be partitioned in the form of $Q \triangleq \begin{bmatrix} Q_1 & Q_2 \\ Q_2^T & Q_3 \end{bmatrix}$ and $W \triangleq \begin{bmatrix} W_1 & W_2 \\ W_4 & W_3 \end{bmatrix}$. There is no loss of generality in assuming that $W_4$ and $W_3$ are invertible. Define the following matrices

$$K \triangleq \begin{bmatrix} I & 0 \\ 0 & W_3^{-1}W_4 \end{bmatrix}, \ \tilde{K} \triangleq diag\begin{bmatrix} K, K, I_q, K \end{bmatrix}, \ \bar{K} \triangleq diag\begin{bmatrix} K, I_q, I_m \end{bmatrix} \tag{D.1}$$

and define

$$\bar{Q} \triangleq \begin{bmatrix} \bar{Q}_1 & \bar{Q}_2 \\ \bar{Q}_2^T & \bar{Q}_3 \end{bmatrix} = K^T Q K \tag{D.2}$$

Performing congruence transformations to (8) by $\bar{K}$ and to (9) by $\tilde{K}$, respectively, and taking into account (5), we obtain (D.3) and (D.4) as follows:

$$\begin{bmatrix} \lambda \bar{Q}_1 & \lambda \bar{Q}_2 & 0 & C_1^T - C_2^T D_f^T \\ * & \lambda \bar{Q}_3 & 0 & -W_4^T W_3^{-T} C_f^T \\ 0 & 0 & (\gamma - \mu)I_q & D_{11}^T - D_{21}^T D_f^T \\ * & * & * & \gamma I_m \end{bmatrix} > 0 \tag{D.3}$$

$$\begin{bmatrix} \bar{Q} - \psi_1 - \psi_1^T & (1 + \lambda \varepsilon/2)\psi_1^T + \varepsilon \psi_3 & \sqrt{\varepsilon}\psi_2 & 0 \\ * & -\bar{Q} & 0 & \sqrt{\varepsilon}\psi_4 \\ * & * & -\mu I & \psi_5 \\ * & * & * & \bar{Q} - \psi_1 - \psi_1^T \end{bmatrix} < 0 \tag{D.4}$$

where

$$\psi_1 = \begin{bmatrix} W_1 & W_2 W_3^{-1} W_4 \\ W_4^T W_3^{-T} W_4 & W_4^T W_3^{-1} W_4 \end{bmatrix}$$

$$\psi_2 = \begin{bmatrix} W_1^T B_1 + W_4^T B_f D_{21} \\ W_4^T W_3^{-T} W_2^T B_1 + W_4^T B_f D_{21} \end{bmatrix}$$

$$\psi_3 = \begin{bmatrix} W_1^T A + W_4^T B_F C_2 & W_4^T A_f W_3^{-1} W_4 \\ W_4^T W_3^{-T} W_2^T A + W_4^T B_f C_2 & W_4^T A_f W_3^{-1} W_4 \end{bmatrix}$$

$$\psi_4 = \begin{bmatrix} G_1^T W_1 & G_1^T W_2 W_3^{-1} W_4 \\ 0 & 0 \end{bmatrix}$$

$$\psi_5 = \begin{bmatrix} G_2^T W_1 & G_2^T W_2 W_3^{-1} W_4 \end{bmatrix}$$

By Defining

$$\begin{aligned} \bar{R} &= W_1 \\ \bar{S} &= W_2 W_3^{-1} W_4 \\ \bar{T} &= W_4^T W_3^{-1} W_4 \\ \begin{bmatrix} \bar{A}_f & \bar{B}_f \\ \bar{C}_f & \bar{D}_f \end{bmatrix} &= \begin{bmatrix} W_4^T & 0 \\ 0 & I \end{bmatrix} \begin{bmatrix} A_f & B_f \\ C_f & D_f \end{bmatrix} \begin{bmatrix} W_3^{-1} W_4 & 0 \\ 0 & I \end{bmatrix} \end{aligned} \tag{D.5}$$

(D.3) and (D.4) are equivalent to (12a) and (12b), respectively.

(b) If there exists a solution to (12a,b), from (D.1) we obtain that

$$\begin{bmatrix} A_f & B_f \\ C_f & D_f \end{bmatrix} = \begin{bmatrix} W_4^{-T} & 0 \\ 0 & I \end{bmatrix} \begin{bmatrix} \bar{A}_f & \bar{B}_f \\ \bar{C}_f & \bar{D}_f \end{bmatrix} \begin{bmatrix} W_4^{-1} W_3 & 0 \\ 0 & I \end{bmatrix} \tag{D.6}$$

Applying (D.6) in the transfer function matrix of the deconvolution filter, $H_{\hat{z}y}(s)$, yields

$$H_{\hat{z}y}(s) = \bar{C}_f W_4^{-1} W_3 \left( sI - W_4^{-T} \bar{A}_f W_4^{-1} W_3 \right)^{-1} W_4^{-T} \bar{B}_f + \bar{D}_f \tag{D.7}$$

After carrying out some transformations, we find that

$$\begin{aligned} H_{\hat{z}y}(s) &= \bar{C}_f \left( sI - W_4^{-1} W_3 W_4^{-T} \bar{A}_f \right)^{-1} W_4^{-1} W_3 W_4^{-T} \bar{B}_f + \bar{D}_f \\ &= \bar{C}_f \left( sI - \bar{T}^{-1} \bar{A}_f \right)^{-1} \bar{T}^{-1} \bar{B}_f + \bar{D}_f \end{aligned} \tag{D.7}$$

which means (13) is established and the proof is completed.

**Appendix E. Proof of Theorem 3**

**Proof.** This theorem can be proved by following similar lines as in the proof of Theorem 3.1 in [5].

According Corollary 2, an admissible improved robust stochastic $L_\infty$-induced filter in the form of (2) exists if for given $\lambda$ and $\varepsilon$, there exist $\bar{Q}(\alpha) > 0$, $\bar{R}(\alpha) \in \Re^{n \times n}$, $\bar{S}(\alpha) \in \Re^{n \times n}$, $\bar{T} \in \Re^{n \times n}$, $\bar{A}_f \in \Re^{n \times n}$, $\bar{B}_f \in \Re^{n \times r}$, $\bar{C}_f \in \Re^{m \times n}$, $\bar{D}_f \in \Re^{m \times r}$, and $\mu > 0$ satisfying (14a,b). Now assume the above matrices to be of the following form:

$$\begin{aligned} \bar{Q}(\alpha) &= \sum_{i=1}^{s} \alpha_i \bar{Q}_i = \sum_{i=1}^{s} \alpha_i \begin{bmatrix} \bar{Q}_{1i} & \bar{Q}_{2i} \\ * & \bar{Q}_{3i} \end{bmatrix}, \\ \bar{R}(\alpha) &= \sum_{i=1}^{s} \alpha_i \bar{Q}_i, \quad \bar{S}(\alpha) = \sum_{i=1}^{s} \alpha_i \bar{S}_i \end{aligned} \tag{E.1}$$

Then, with (E.1) we can easily prove that (14a) holds if and only if (15a) holds. In addition, with (E.1) it is not difficult to rewrite $\Xi\left(\bar{Q}(\alpha), \bar{R}(\alpha), \bar{S}(\alpha), \bar{T}, \bar{A}_f, \bar{B}_f\right)$ as

$$\Xi\left(\bar{Q}(\alpha), \bar{R}(\alpha), \bar{S}(\alpha), \bar{T}, \bar{A}_f, \bar{B}_f\right) = \sum_{j=1}^{s} \sum_{i=1}^{s} \alpha_i \alpha_j \Xi_{ij} = \sum_{i=1}^{s} \alpha_i^2 \Xi_{ii} + \sum_{i=1}^{s-1} \sum_{j=i+1}^{s} \alpha_i \alpha_j \left( \Xi_{ij} + \Xi_{ji} \right) \tag{E.2}$$

where $\Xi_{ij}$ is defined in (17) and (18). Then (15b) and (15c) guarantee $\Xi\left(\bar{Q}(\alpha), \bar{R}(\alpha), \bar{S}(\alpha), \bar{T}, \bar{A}_f, \bar{B}_f\right) < 0$, and the proof is completed.